\documentclass[twocolumn,aps,prb,superscriptaddress,10pt]{revtex4-2}

\usepackage{graphicx}
\usepackage{amsmath}
\usepackage{siunitx}
\usepackage{units}

\begin{document}

\title{Low charge-noise nitrogen-vacancy centers in diamond created using laser writing with a solid-immersion lens}

\author{V.~Yurgens} 
\email{viktoria.yurgens@unibas.ch}
\author{J.~A.~Zuber} 
\author{S.~Fl\aa gan} 
\author{M.~De~Luca} 
\author{B.~J.~Shields}
\author{I.~Zardo} 
\author{P.~Maletinsky} 
\author{R.~J.~Warburton}
\affiliation{Department of Physics, University of Basel, CH-4056 Basel, Switzerland}
\author{T.~Jakubczyk} 
\affiliation{Department of Physics, University of Basel, CH-4056 Basel, Switzerland}
\affiliation{Faculty of Physics, University of Warsaw, 02-093 Warsaw, Poland} 

\begin{abstract}
We report on pulsed-laser induced generation of nitrogen-vacancy (NV) centers in diamond facilitated by a solid-immersion lens (SIL). The SIL enables laser writing at energies as low as \unit[5.8]{nJ} per pulse and allows vacancies to be formed close to a diamond surface without inducing surface graphitization. We operate in the previously unexplored regime where lattice vacancies are created following tunneling breakdown rather than multiphoton ionization. We present three samples in which NV-center arrays were laser-written at distances between \SI{\sim 1}{\micro\meter} and \SI{40}{\micro\meter} from a diamond surface, all presenting narrow distributions of optical linewidths with means between \unit[62.1]{MHz} and \unit[74.5]{MHz}. The linewidths include the effect of long-term spectral diffusion induced by a \unit[532]{nm} repump laser for charge-state stabilization, thereby emphasizing the particularly low charge-noise environment of the created color centers. Such high-quality NV centers are excellent candidates for practical applications employing two-photon quantum interference with separate NV centers. Finally, we propose a model for disentangling power broadening from inhomogeneous broadening in the NV center optical linewidth.
\end{abstract}

\maketitle

\section{Introduction}
The negatively charged nitrogen-vacancy (NV) center in diamond is among the most promising solid-state systems implementations of a quantum bit \cite{Jelezko2004a,Jelezko2004b}. Its applications include sensing \cite{Maze2008,Rondin2014} and quantum communication, with progress demonstrated in spin-photon \cite{Togan2010} and long-distance spin-spin entanglement \cite{Hensen2015}. However, interconnecting many NV centers for large-scale quantum networks suffers from the low generation rate of indistinguishable photons from individual NV centers \cite{Hensen2015}. 

Fabrication of any diamond-based photonic device requires structuring of the diamond on a sub-micron scale. An example is the open Fabry-Perot microcavity -- a platform enabling easy optical access, mode-matching and \emph{in situ} tuning of the cavity resonance to an emitter \cite{Barbour2011b,Albrecht2013,Johnson2015,Kaupp2016}. In this case, bulk diamond is thinned down to a few-micron-thick membrane, a minimally invasive fabrication \cite{Janitz2015,Bogdanovic2017,Riedel2017c,Janitz2020,Heupel2020,Riedel2020}. The technique is a promising route to enhance radically the generation rate of indistinguishable photons from individual NV centers. However, even such minimal processing has so far resulted in degradation of the NV centers' optical quality, manifested in a large spectral diffusion \cite{Riedel2017c,Ruf2019a}. The random spectral fluctuations in the zero-phonon line (ZPL) frequency are caused by charge noise present in the structured crystal: NV centers exhibit a large change in the static dipole moment between the ground- and excited state \cite{Maze2011}, making the ZPL emission frequency strongly dependent on the local electric field \cite{Schmidgall2018}. The electric field in turn depends on the spatial configuration and occupation of the surrounding charge traps in the form of defects and impurities. Methods of NV center creation that do not result in the formation of parasitic defects and impurities are therefore desired.

The creation of an NV center typically involves displacing a carbon atom from its lattice site to create a vacancy. This occurs above a threshold energy of \SI{35}{eV} \cite{Bourgoin1976}. While proximity to the surface is required for photonic applications, it is at the same time desirable to form the NV centers at a depth of at least tens of nanometers from the diamond surface in order to minimize the impact of the surface charge- and magnetic-noise on the NV centers' spin- and optical coherence. The challenge is therefore to provide this energy inside the crystal lattice. Electrons or ions with kinetic energies far above keVs readily provide such an energy to depths from a few nanometers to tens of micrometers and above in the case of electrons \cite{Campbell2000}. However, most of this energy is released via collisions on the particle's trajectory, leaving extended damage and presumably hard-to-anneal vacancy complexes. A widely employed technique to create NV centers at precisely controlled depths consists of nitrogen-ion implantation followed by high-temperature annealing \cite{Chu2014a}. The technique is well suited for creating NV centers for sensing applications \cite{Pezzagna2011,Appell2016}, but has recently been shown to create NV populations where approximately half of the NV centers show spectral diffusion above a few GHz \cite{VanDam2019,Kasperczyk2020}. In addition, the NV centers formed from the implanted nitrogen ions have on average much poorer optical quality than their counterparts formed from the native nitrogen \cite{VanDam2019,Kasperczyk2020}, suggesting that vicinity to the fabrication-related damage plays a crucial role for the spectral diffusion. 

Typically, the inhomogeneously-broadened linewidth distribution in unprocessed bulk samples irradiated by ions or electrons is characterized by linewidths on the order of \unit[100]{MHz} \cite{Ruf2019a,VanDam2019,Kasperczyk2020}, disregarding half of the population with above-GHz linewidths for the nitrogen-implanted samples. After thinning the samples down to less than a micron by reactive ion etching, most of the observed NVs are characterized by \unit[$>1$]{GHz} linewidths \cite{Riedel2017c}. This increase in linewidth is detrimental for photon indistinguishability \cite{Bernien2012,Sipahigil2012}. Good two-photon interference is only possible when the inhomogeneous broadening is small compared to the homogeneous linewidth. For an NV center in bulk diamond, the lifetime limit sets the smallest possible homogeneous broadening of just \unit[$\sim$13]{MHz} \cite{Tamarat2006}, far below typical inhomogeneous linewidths. However, in a resonant microcavity, the Purcell effect strongly reduces the lifetime, thereby increasing the lifetime limit. Nevertheless, in current devices, the inhomogeneous broadening is still higher than the Purcell-enhanced homogeneous linewidth \cite{Riedel2017c}. Radical progress in improving the emitter properties or microcavity properties, ideally both, is therefore required. 

A recent study demonstrated progress in obtaining NV centers with linewidths of approximately \unit[200]{MHz} in \SI{3.8}{\micro\metre}-thick diamond membranes \cite{Ruf2019a}. In even thinner structures, linewidths down to \unit[250]{MHz} were reported with a new approach based on implanting nitrogen after micro-structuring \cite{Kasperczyk2020} and also in samples structured with a slow etching procedure \cite{Lekavicius2019}. To the best of our knowledge, narrower linewidths than these have never been reported in micron-thick diamond. The origin of the decrease in the NV centers' optical quality is potentially related to the damage caused to the crystal lattice during implantation and etching, but the microscopics are not well understood. The lack of optically coherent NV centers in structured diamond clearly points to the need for improved NV creation methods.

Laser writing has emerged as a promising technique for creating deterministically positioned NV centers in diamond \cite{Chen2017b} and other color centers in various wide-bandgap materials \cite{Buividas2015,Hou2017,Castelletto2018,Umar2018,Chen2019b}. NV centers created through laser writing in ultrapure diamond are characterized by long spin coherence times, low spectral noise and high charge stability \cite{Chen2017b}. Crucially, their fabrication induces minimum damage to the diamond lattice, potentially reducing the charge noise in thin diamond membranes. These NV centers are formed from native nitrogen atoms and implantation is avoided completely, meaning that the lattice is damaged only locally in the laser focus. 

Initial studies demonstrated laser writing of NV centers using an oil-immersion lens in combination with wavefront correction \cite{Chen2017b, Chen2019}. Here, we report on a different approach employing a standard air objective together with a solid-immersion lens (SIL), giving a high numerical aperture (NA) together with minimized aberrations, yielding a diffraction-limited laser focal volume inside diamond. This technique not only enables creation of NV centers with low pulse energies, but also makes it possible to laser-write in close proximity to the diamond surface owing to the decreased laser intensity at the diamond-air interface. We demonstrate the high quality of the NV centers created with this method, with a record-low noise measured in the presence of a charge-state reinitialization pulse, and determine separately the inhomogeneous and power-broadening contributions to the ZPL linewidths.

\begin{figure}[t!]
	\centering
		\includegraphics[width=1\columnwidth]{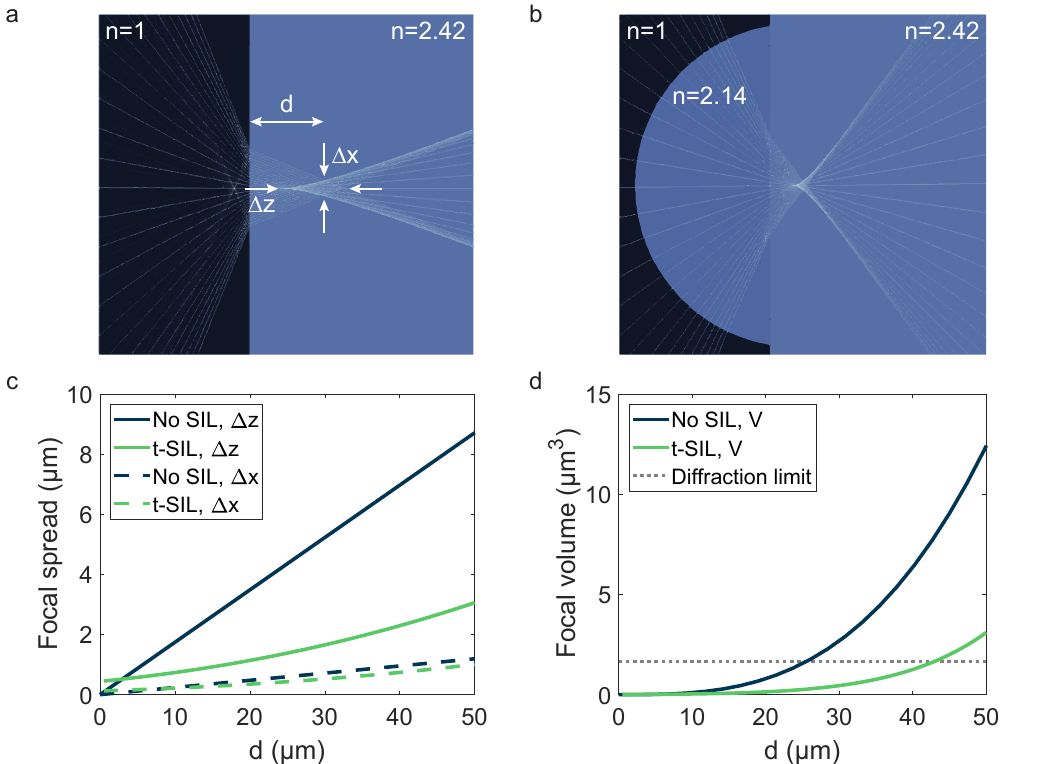}
	\caption{Geometrical optics simulation of the spread of rays on focusing through an air-diamond interface. (a) Spherical aberration for rays passing from air (left) to diamond (right) without corrective optics. $\Delta x$ is the in-plane focal spread and $\Delta z$ is the focal spread along the optical axis; $d$ is the distance from the point of minimum $\Delta x$ to the diamond surface. (b) Spherical aberration for rays passing from air to diamond through a truncated hemispherical solid-immersion lens (t-SIL). (a,b) were calculated using the tool from Ref.~\citenum{rayoptic}. (c) Calculated focal spread in plane (solid lines) and along the optical axis (dashed lines) for the cases without a SIL and with a t-SIL on passing through an air-diamond interface, as a function of the focusing depth $d$. (d) Calculated focal volume versus $d$ without and with a t-SIL. (c) and (d) take a t-SIL truncation of \SI{30}{\micro\meter} and an NA of 0.6. The geometrical optics simulations ignore diffraction which determines the focal volume when aberrations are absent. In (d), the diffraction limit is estimated as $(\Delta x)^2 \Delta z$ with $\Delta x = 1.2 \lambda /(n {\rm NA})$ and $\Delta z=2.0 \lambda/(n {\rm NA}^2)$, with $n=2.42$.}
	\label{fig:SIL}
\end{figure}

\section{Laser writing of NV centers}

Laser writing of NV centers is based on irradiation of a diamond sample with single high-energy femtosecond pulses. The laser pulses create vacancies; NV centers form in a subsequent thermal annealing step in which the mobile vacancies combine with nitrogen impurities in the diamond. Photo-induced damage in a transparent wide-bandgap material such as diamond requires the generation of excited free electrons through nonlinear mechanisms. This optical breakdown can result from tunneling or multi-photon absorption followed by avalanche ionization in the volume of the tightly focused laser beam \cite{Schaffer2001,Mao2004, Gattass2008}. These processes displace single atoms locally and can create vacancies. In diamond, vacancy creation can be identified by the vacancy's spectral signature GR1 \cite{Clark1973}. The dominant mechanism for the creation of the energetic seed electrons responsible for the optical breakdown can be determined using the Keldysh parameter \cite{Keldysh1964}
\begin{equation}
    \gamma = \frac{\omega}{e}\sqrt{\frac{m_{\rm eff}cn\epsilon_0 E_g}{I}}
\end{equation}
where $\omega$ is the angular frequency of the laser, $I$ the laser intensity, $m_{\rm eff}$ the electron effective mass, $c$ the speed of light in vacuum, $n$ the linear refractive index of the material, $\epsilon_0$ the vacuum permittivity, and $E_g$ the direct bandgap of the material. For $\gamma<1$, tunneling breakdown is the primary process behind the creation of seed electrons, while for $\gamma>1$, multi-photon ionization dominates. In the multi-photon ionization regime, owing to the strong nonlinearity of the process \cite{Chen2017b}, the volume in which the atom displacement takes place can be much smaller than size of the focal spot. However, during the annealing step, the created vacancies diffuse and this process limits the NV placement accuracy, offering a precision down to hundreds of nanometers in initial reports \cite{Chen2017b, Chen2019}. The laser-written NV centers exhibit a robust charge state with little need for charge-state repump \cite{Chen2017b}. However, the question remains if the good optical quality is maintained after processing of the diamond.

To minimize the impact of the laser writing process on the crystal quality, the volume of the laser focus needs to be reduced by maximizing the NA of the focusing optics. However, spherical aberration induced at the air-diamond interface (with a refractive index of $n=1$ and $n=2.42$, respectively) also increases with increasing NA. We resolve this dilemma by using a truncated hemispherical cubic zirconia SIL (t-SIL) with a refractive index of $n=2.14$, which both maximizes the NA and reduces the spherical aberrations (Fig.\,\ref{fig:SIL}(a),(b)).

An absolutely crucial problem is that the energy of the ultra-short laser pulses required to create vacancies in bulk diamond typically exceeds the threshold at which the diamond surface is degraded \cite{Konov2012,Chen2017b,Kononenko2017b}. Creating vacancies in bulk diamond with laser writing is therefore only possible if the laser focus is spread over a significant area at the surface. To investigate this, we consider the case of imaging a laser with wavelength 800 nm with an objective lens of NA=0.6 in Fig.\,\ref{fig:SIL}. In air, the extent of the focal spot along the optical axis, $\Delta z$, is in the best case limited by diffraction to a value of approximately \SI{3.9}{\micro\meter}. This extent will be preserved approximately on focusing a few microns below the diamond surface. The large $\Delta z$ is problematic: the intensity at the surface is large, and as a result, surface graphitization will occur before vacancy creation within the diamond. Focusing at a larger depth does not solve this problem. Geometrical optics shows that the spherical aberrations cause $\Delta z$ to increase with depth $d$ (Fig.\,\ref{fig:SIL}(c)) -- the focal spot becomes so elongated along $z$ that surface graphitization is likely to remain a problem. This corresponds to our experience -- without a t-SIL, surface graphitization occurred before vacancy creation. In the limit that the t-SIL is made of diamond, spherical-aberration-free imaging is achieved at a depth $d$ which matches the truncation $t$. Furthermore, the diffraction limit reduces simply because the relevant wavelength is the wavelength in the diamond, $\lambda/n$. For NA$=0.64$, $\Delta z$ reduces to \SI{1.8}{\micro\meter}. These considerations suggest that vacancy creation is possible at depths starting at about \SI{2}{\micro\meter} from the diamond surface and it motivates the use of the t-SIL. In practice, the t-SIL has a slightly lower refractive index than diamond (such that a spherical aberration is introduced at the t-SIL--diamond interface) and it is impractical to use a different t-SIL for each depth. We find however that a single t-SIL gives good imaging out to a depth up to \SI{50}{\micro\meter} (Fig.\,\ref{fig:SIL}(c),(d)). The main improvement over imaging without the t-SIL is the reduction in $\Delta z$, crucial in order to keep the intensity at the surface small.

\begin{figure}[t!]
	\centering
		\includegraphics[width=1\columnwidth]{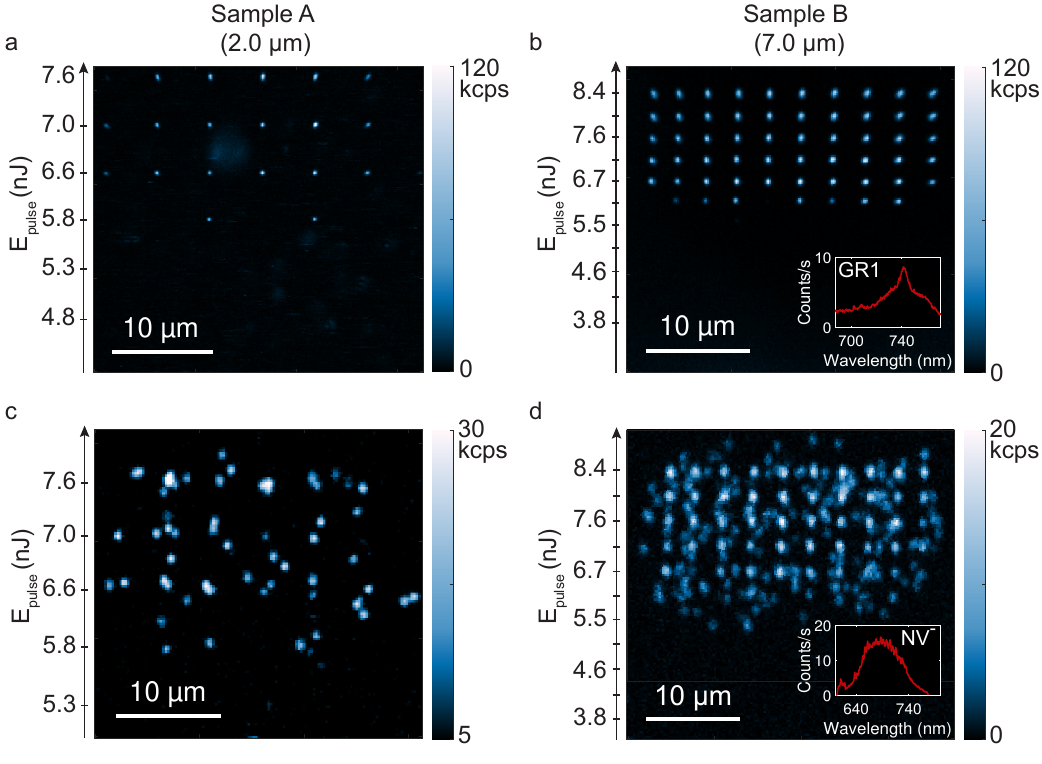}
	\caption{(a,b) Confocal scans of samples A and B (indicating their respective NV creation depths and pulse energies) showing GR1 photoluminescence (inset in b) at room temperature before annealing. The scans were performed through a t-SIL. (c,d) Confocal scans of samples A and B, showing NV$^{-}$ photoluminescence (inset in d) at room temperature after annealing to 1100$^\circ$C.}
	\label{fig:confocals}
\end{figure}

An additional advantage is that the t-SIL improves the photon collection efficiency and thereby the sensitivity to the weak signals emitted by the created vacancies by reducing the number of photons lost due to total internal reflection at the diamond-air interface \cite{Serrels2008a,Serrels2008b,Riedel2014}. Moreover, the implementation of a t-SIL is both cost-effective and easily implemented. A limitation is perhaps the writing area -- excellent imaging is achieved only close to the center of the t-SIL. However, a writing area of approximately 25-by-25$\,$\SI{}{\micro\meter}$^2$ is available for a t-SIL with a diameter of \unit[0.5]{mm}, sufficiently large for many purposes.

We perform laser writing in a room temperature home-built confocal microscope with a separate injection path for the femtosecond pulsed laser. A \unit[532]{nm} Nd:YAG laser is used for non-resonant excitation of the NV centers. Photoluminescence (PL) is filtered by a longpass filter (Semrock, 594 nm RazorEdge) and collected by an APD (Excelitas, SPCM-AQRH-15-FC) or a liquid nitrogen-cooled CCD camera coupled to a grating spectrometer (Princeton Instruments). We use a Spectra Physics Spirit 1030-70 femtosecond ytterbium-doped fiber laser together with a Spirit-NOPA-2H non-collinear optical parametric amplifier, which together create pulses at a wavelength of \unit[800]{nm} with a duration of approximately \unit[35]{fs}. A half-wave plate and a Brewster-angle polarizer are used for tuning the pulse energies. The flat side of a t-SIL (radius \SI{500}{\micro\meter}, truncation between 0 and \SI{50}{\micro\meter}) is polished using a polishing suspension containing \unit[50]{nm}-sized alumina particles. The t-SILs are then placed on the diamond samples together with an optical coupling gel in an attempt to reduce aberrations resulting from interface imperfections. We focus the \unit[800]{nm} laser with a standard air objective (Olympus, MPLFLN100x, NA=0.9 or Olympus, LCPLFLN100xLCD, NA=0.85). The resultant NA including the effect of the t-SIL is about 1.8, slightly higher than that used in Ref.~\citenum{Chen2017b}. We note however that the full-width-at-half-maximum (FWHM) of the beam is about the same as the input aperture of the objective, reducing the contribution of the rays at the highest angle. The performance of the optical setup including the t-SIL was determined by confocal imaging of the emission from single, preexisting, shallow NV centers. The FWHM of a single bright spot was measured at \unit[$125\pm3$]{nm}. This value is just above the theoretical minimum of \unit[121.1]{nm}, as given by $\frac{0.52}{\text{NA}}\frac{\lambda_1\lambda_2}{\sqrt{\lambda_1^2+\lambda_2^2}}$ for the confocal configuration (using \unit[$\lambda_1 = 532$]{nm} and \unit[$\lambda_2 = 700$]{nm} for the excitation and main NV phonon-sideband emission wavelength, respectively, and taking NA$=0.85\cdot2.14$). This close agreement between the measured spot size and the diffraction limit demonstrates that the optical components, notably the objective and the t-SIL, do not introduce unwanted aberrations.

We follow the experimental procedure described in Ref.~\citenum{Chen2017b} to create arrays of vacancies in bulk electronic-grade diamond samples (\SI{40}{\micro\meter} thick, [N]$<$\unit[5]{ppb}, Element Six). We use single pulses at increasing pulse energies (\unit[3.8-35.8]{nJ}) and subsequently anneal the diamond samples at 1100$^\circ$C to create NV centers. Surface markers for locating the laser-written arrays after annealing are made through graphitization of the diamond surface by increasing the pulsing frequency of the femtosecond laser to \unit[100]{kHz}, removing the t-SIL, and increasing the energy to a value above the graphitization threshold of the surface (typically around \unit[5]{nJ}).

Results on two diamond samples are shown in Fig.\,\ref{fig:confocals}, with arrays of vacancies (a,b) and NV centers (c,d), respectively, created with increasing pulse energies from the bottom of the arrays. The scans were recorded at room temperature. The arrays were made at a depth of \SI{2.0}{\micro\meter} and \SI{7.0}{\micro\meter} below the diamond surface, respectively, where we note that the former demonstrates that our method can be used to create NV centers in close proximity to the diamond surface without inducing graphitization. Vacancy creation at shallower depths was not possible without damaging the diamond surface. This observation aligns with the analysis of the dimensions of the focal spot. For sample A, pulse energies of between \unit[4.8]{nJ} and \unit[7.6]{nJ} were used, with visible GR1 photoluminescence (inset Fig.\,\ref{fig:confocals}(b), \unit[100]{s} integration time) appearing for \unit[5.8]{nJ} per pulse; for sample B, pulse energies between \unit[3.8]{nJ} and \unit[8.4]{nJ} were used, with visible GR1 photoluminescence appearing for \unit[6.1]{nJ} per pulse. All pulse energies refer to values at the output of the microscope objective. Both samples were annealed in vacuum for three hours at 1100$^\circ$C, after which the spectral signatures of negatively charged NV centers (inset Fig.\,\ref{fig:confocals}(d), \unit[100]{s} integration time) were observed.

Calculation of the Keldysh parameter $\gamma$ for these vacancy-creation threshold energies, with the specified pulse duration and focal volume, gives $\gamma\ll1$, meaning that we operate in a regime where tunneling breakdown is dominant over multi-photon ionization, in contrast to previous work \cite{Chen2017b}. Tuning of the vacancy creation regime could be carried out by changing the laser writing wavelength and increasing the pulse length, but with the latter shrinking the window for NV center creation due to a decrease of the graphitization threshold \cite{Kurita2018}.

\begin{figure}[t!]
	\centering
		\includegraphics[width=1\columnwidth]{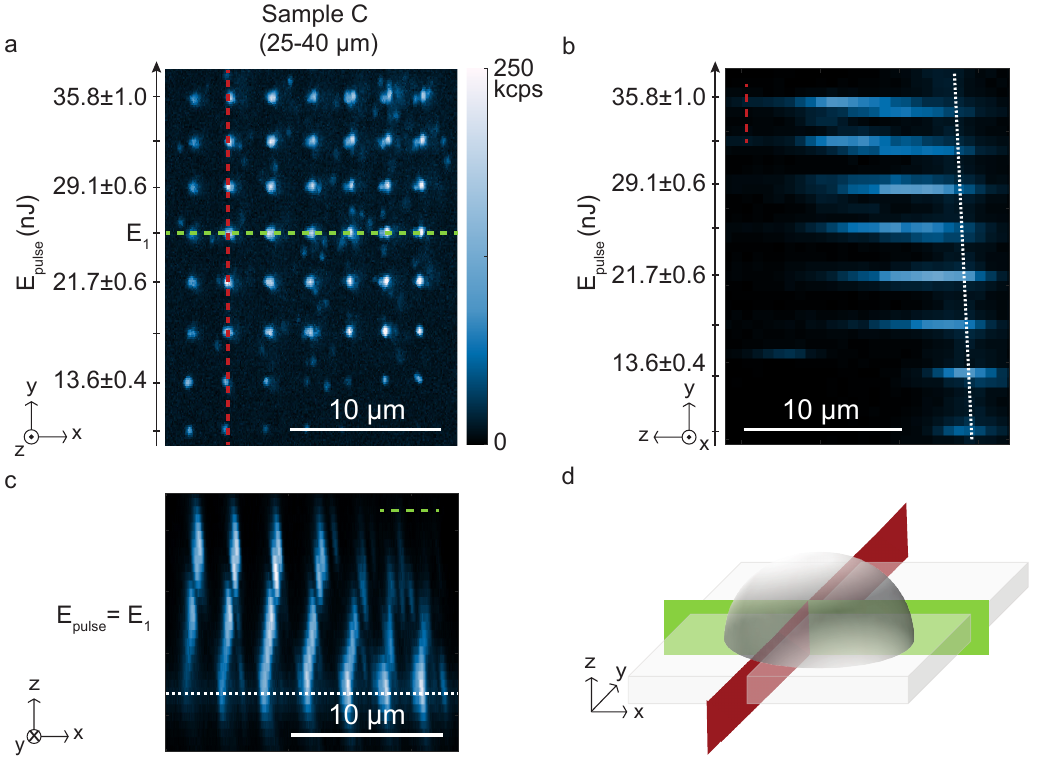}
	\caption{(a) Confocal scan of NV centers in sample C, laser-written in an inverse geometry where laser pulses were focused close to the bottom surface of the diamond. Each consecutive spot along a line was laser written at a reduced distance from the surface compared to its neighbor. The colorbar applies to a-c. Some of the spots contain multiple NV centers -- this accounts for the changes in brightness from spot to spot. (b) Cross-section in the plane indicated by the red dashed-line in (a). The white dashed-line indicates the bottom diamond surface, determined by the position at maximum intensity of the reflected excitation laser. (c) Cross-section in the plane indicated by the green dashed-line in (a), for a pulse energy E$_1=$\unit[25.7$\pm$0.5]{nJ}. The white dashed-line indicates the bottom diamond surface. (d) Schematic of the two cross-sections through the diamond, with a t-SIL placed on top.}
	\label{fig:00019C}
\end{figure}

Sample B shows a clear array-like NV pattern after annealing, where the main array points contain multiple NV centers, as demonstrated by the photon count in the confocal scans and by the existence of more than two lines in subsequent photoluminescence excitation (PLE) measurements. The diffusion of vacancies during annealing is also clearly visible -- single NV centers form up to several hundreds of nanometers from the array spots, similar to the results of previous experiments \cite{Hu2002,Onoda2017,Chen2017b}. Lower pulse energies and a larger spatial separation between the focal spots were therefore used for sample A in order to obtain predominantly spots with single- rather than multiple NV centers. 

Vacancy photoluminescence after laser writing was observed both with the t-SIL and after displacing the t-SIL from the laser-processed area. If no t-SIL was used during the laser-writing process, neither graphitization nor GR1 photoluminescence could be detected in the bulk of the sample up to a pulse energy of \unit[52]{nJ} (highest available in our experimental configuration) and a continuous exposure with a pulse repetition rate of \unit[1]{MHz} with durations up to tens of seconds. Similar observations were made after exposures performed with an t-SIL but without using the index-matching gel. Also after annealing there was no NV photoluminescence in areas exposed without a t-SIL. These observations indicate that the use of a t-SIL dramatically reduces the threshold for vacancy-generation and can act as the key element enabling laser-induced NV formation using fs laser sources.  

A third sample, sample C, was patterned in a way to create NV centers as close to the diamond surface as possible without inducing visible damage \cite{Liu2013}. PL images of the resulting array of NV centers are shown in Fig.\,\ref{fig:00019C}(a-c). The increased energy density on the top diamond surface upon focusing the laser less than \SI{\sim 2}{\micro\meter} from it typically resulted in graphitization of the diamond surface or damage of the t-SIL. For this reason, an inverse geometry was chosen. In this geometry, the laser pulses were applied through the t-SIL and diamond, but focused close to the diamond back surface, the one further from the t-SIL. Several NV arrays were patterned at different depths in the diamond in this configuration. An objective lens with NA=0.85 and correction ring was used for this study to further minimize the spherical aberration; the correction ring was set to maximize the laser focus intensity at the bottom surface of the sample.

NV centers could be created in the full range of pulse energy in Fig.\,\ref{fig:00019C}(a), from \unit[$7.6\pm 0.4$]{nJ} to \unit[$35.8 \pm 1.0$]{nJ}. Even at the highest pulse energies there were no signs of graphitization. This energy window is remarkably wide compared to previous studies \cite{Chen2017b, Chen2019}. We propose that this is a key consequence and advantage of NV center laser writing in the tunneling regime instead of the multi-photon ionization regime. In order to test NV writing at various depths and in order to find a possible distance limit from the surface for NV creation, each array additionally included a smaller depth variation of \SI{0.4}{\micro\meter} between each laser writing spot within a row. This resulted in NV centers covering distances from \SI{\sim 1}{\micro\meter} up to \SI{15}{\micro\meter} from the bottom diamond surface, corresponding to laser writing depths of between \SI{25}{\micro\meter} and \SI{40}{\micro\meter} from the top surface. Fig.\,\ref{fig:00019C}(b),(c) show cross sections through the diamond to illustrate the resulting slightly tilted layers. In Fig.\,\ref{fig:00019C}(b) it is apparent that a lower pulse energy was needed to create the NV centers closest to the surface, as compared to writing NV centers further inside the bulk. The reason for this is presently unknown. Fig.\,\ref{fig:00019C}(d) illustrates the cross sections and the general t-SIL-on-diamond geometry.

\begin{figure}[t!]
	\centering
    \includegraphics[width=1\columnwidth]{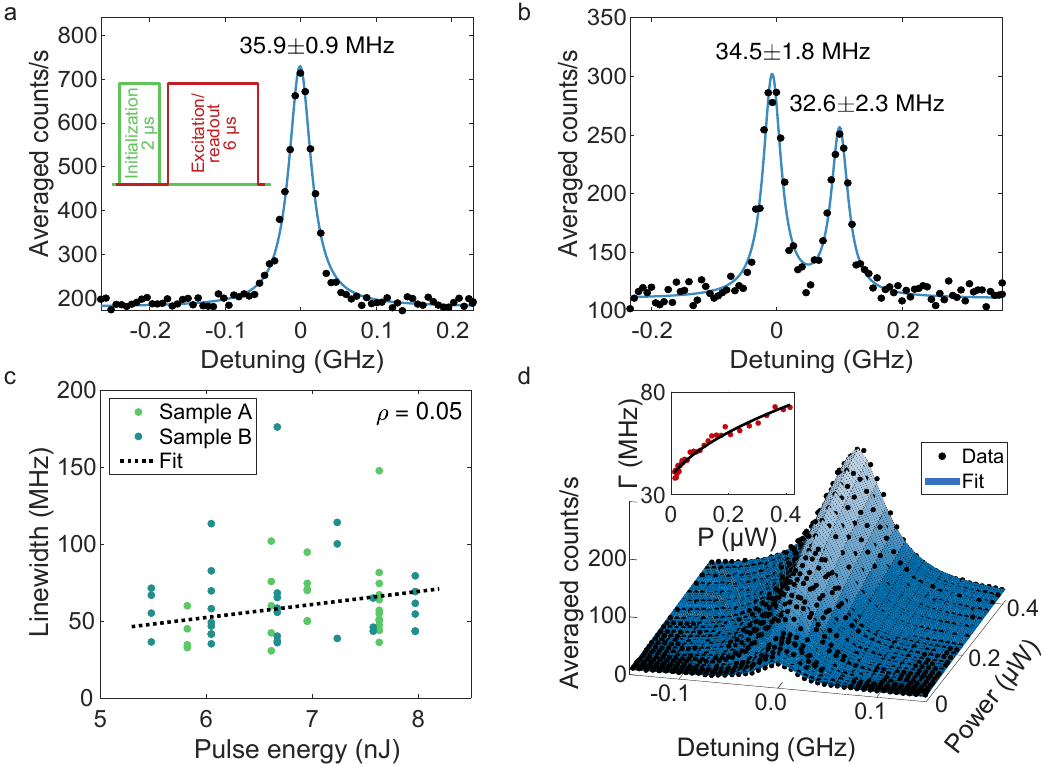}
	\caption{(a) Optical linewidth measured on an NV in sample A. Zero detuning corresponds to a ZPL frequency of \unit[470.494]{THz}. The inset shows the measurement sequence for the photoluminescence excitation measurement. (b) Linewidth measured on an NV center in sample B. Zero detuning corresponds to a ZPL frequency of \unit[470.509]{THz}. (c) Dependence of the linewidths in the two samples on the fs laser pulse energy used in the laser writing process. Each data point corresponds to a separate NV center, except for a few cases where two lines were measured for one NV center. The dotted line is a linear fit with a Pearson's $\rho=0.05$ indicating little to no correlation between linewidth and pulse energy. (d) Resonant power dependence of the ZPL of an NV in sample A, fitted according to equation~\eqref{eq:fullfit}. The inset shows the dependence of the extracted FWHM linewidth on power, fitted following equation~\eqref{eq:LW_power_fit} and demonstrating clear power broadening.}
	\label{fig:LWs}
\end{figure}

\section{Linewidth characterization}

We characterize the ZPL linewidths through PLE scans in a liquid helium bath cryostat. The NV centers are resonantly excited using a tunable narrow-linewidth external cavity diode laser (New Focus Velocity TLB-6704) using powers between \unit[15]{nJ} and \unit[450]{nJ} and the linewidths are determined by sweeping the excitation frequency over the ZPL resonance while recording the photons emitted into the phonon sideband. Each line is measured at the lowest possible power which gives a good signal-to-noise ratio in order to have as small power broadening as possible. Between the periods of excitation and readout, a negative charge-state repump is carried out with a \unit[532]{nm} Nd:YAG-laser with an average power of \unit[0.6]{mW}. The full cycle of repump followed by resonant excitation and readout is repeated at a frequency of \unit[100]{kHz}, with a 21-to-68 green-to-red laser duty cycle. The inset in Fig.\,\ref{fig:LWs}(a) shows the PLE measurement sequence.

Two PLE scans are presented in Fig.\,\ref{fig:LWs}(a),(b), showing ZPL resonances with Lorentzian lineshapes with FWHM linewidths of \unit[$35.9 \pm 0.9$]{MHz}, \unit[$34.5 \pm 1.8$]{MHz} and \unit[$32.6 \pm 2.3$]{MHz}, close to the lifetime-limited linewidth of \unit[13]{MHz} measured for a native NV center in a bulk natural diamond sample \cite{Tamarat2006}. (In the low-power regime, a Gaussian fits the PLE data only slightly better than a Lorentzian: average mean-standard-errors 0.014 and 0.017, respectively. We use here a Lorentzian for simplicity.) We note that Ref.~\citenum{Tamarat2006} measured linewidths between the repump pulses and thereby excluded a major source of inhomogeneous broadening, charge reconfiguration of the environment on resetting the NV center charge. In our case, we integrate over many repump cycles, thereby including all the contributions to the inhomogeneous broadening. We note also that fitting the lines measured at low power (where inhomogeneous broadening dominates) with Gaussian functions rather than Lorenztians yielded an average difference in FWHMs of only 1.5\% for the three samples, further justifying the use of a Lorentzian function to describe the inhomogeneous broadening. Fig.\,\ref{fig:LWs}(c) shows the dependence of the linewidths on the femtosecond laser pulse energy, where the dotted line is a linear fit to the data. A Pearson's $\rho=0.05$ for the fit indicates no significant correlation between ZPL linewidth and pulse energy, in contrast to Ref.~\citenum{Chen2017b}.

Fig.\,\ref{fig:LWs}(b) shows a splitting between the resonances of just \unit[109]{MHz}, which indicates a low strain level in the surroundings of the specific NV. (Here, we make the reasonable assumption that the peaks correspond to the E$_x$- and E$_y$ transitions of the same color center.) In general, the three laser-written samples demonstrate several doublets with exceptionally low peak splittings down to \unit[81]{MHz}, but a more systematic study is required in order to obtain a definite conclusion on the strain levels of laser-written NV centers and to exclude the possibility that the two lines originate from separate NVs (however, such coincidence in spatial location, linewidth and ZPL energy between different NVs is highly unlikely). The absolute ZPL frequency has a distribution centered at 470.49 THz with a standard deviation of 14 GHz, similar to the distribution in Ref.~\citenum{VanDam2019}.

To disentangle inhomogeneous broadening of the linewidths ($\Gamma_{\rm in}$) from power broadening characterized by the Rabi coupling $\Omega$, we perform a systematic study of the linewidth as a function of resonant power. To simplify the analysis we assume a Lorentzian spectral diffusion shape characterized by FWHM $\Gamma_{\rm in}$: the probability of the emitter frequency being equal to $f^*$ is given by the (normalized) Lorentzian function $L(f^*-f_0, \Gamma_{\rm in})$ where $f_0$ is the average emitter frequency. The occupation of the excited level of a driven two-level system with radiative decay rate $\gamma$ is given by \cite{Loudon2000}
\begin{equation}
    \rho_{22} = \frac{\big(\frac{1}{2}\Omega\big)^2}{4\pi^2(f-f^*)^2+\big(\frac{1}{2}\gamma\big)^2+\frac{1}{2}\Omega^2}
\end{equation}
where $\Omega = \sqrt{c\cdot P}$, with $P$ the excitation power and $c$ an effective coupling strength which depends on the laser's focal volume, laser incoupling efficiency and the NV center's dipole moment and orientation. The experimentally measured line shapes can be described by a convolution of $L$ with $\rho_{22}$, yielding an expression for the counts as a function of frequency in a PLE measurement:
\begin{equation}
    \label{eq:fullfit}
    C(f) = \frac{A}{4\pi} \frac{\Omega^2}{\sqrt{\gamma^2+2\Omega^2}} \frac{\frac{1}{2}\Gamma}{(f-f_0)^2 + \big(\frac{1}{2}\Gamma\big)^2}
\end{equation}
where $A$ depends on the setup's collection efficiency and the average NV photon emission rate, the ''dead'' time during the repump pulse and the time spent in the NV$^0$ charge state, as well as the time spent in spin states that are not cycled with the resonant laser \cite{Chu2015}. The FWHM linewidth is
\begin{equation}
    \label{eq:LW_power_fit}
    \Gamma = \Gamma_{\rm in} + \frac{\sqrt{\gamma^2+2\Omega^2}}{2\pi}
\end{equation}
where we assume \unit[$\gamma =13$]{MHz}.

\begin{figure}
	\centering
		\includegraphics[width=1\columnwidth]{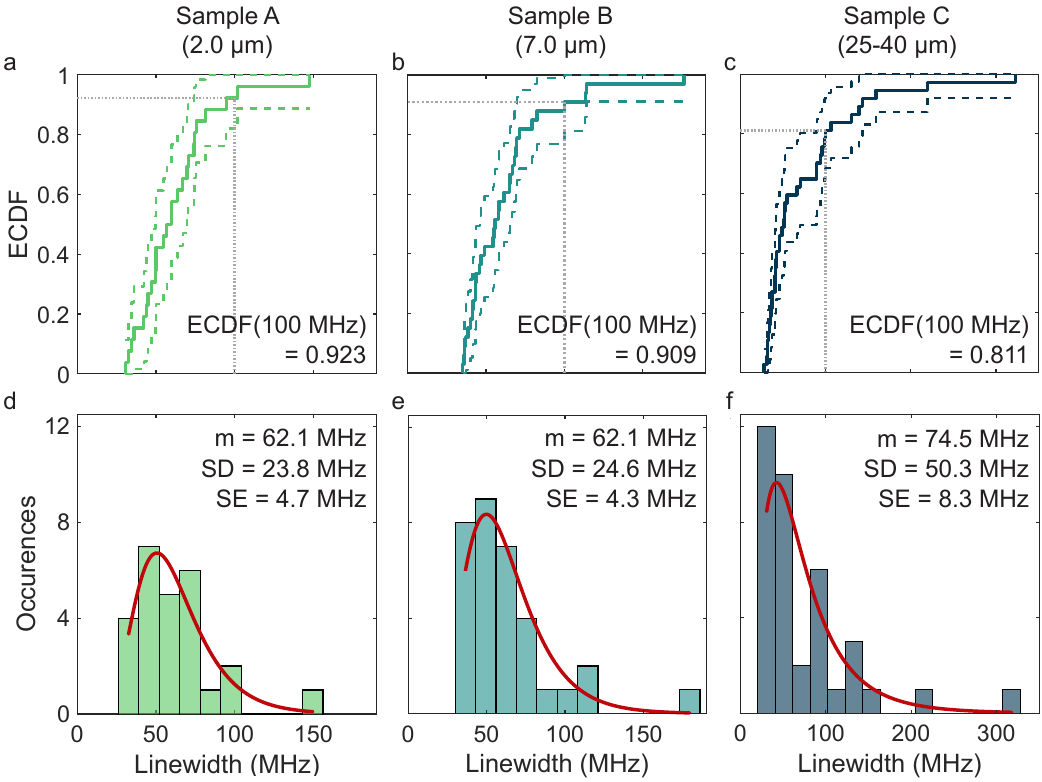}
	\caption{Linewidth statistics of sample A (a,d), B (b,e), and C (c,f), with empirical cumulative distribution functions (ECDF, a-c) with 95\% confidence intervals and linewidth histograms with log-normal fits (d-f). The values of the ECDFs for \unit[100]{MHz} as well as the means (m), standard deviations (SD) and the standard errors (SE) of the mean are specified.}
	\label{fig:stats}
\end{figure}

Fig.\,\ref{fig:LWs}(d) shows a PLE measurement as a function of the resonant excitation power and the data can be fitted well with Eq.\,\eqref{eq:fullfit}. Power broadening is clearly visible. The inset shows the extracted FWHM linewidth as a function of power, fitted with the linewidth described by Eq.\,\eqref{eq:LW_power_fit} (solid black line) as well as with a linewidth given only by the second term of the equation (dashed grey line). The full model gives a value of \unit[$c = (1.0 \pm 0.1) \cdot 10^5$]{MHz$^2$/}\SI{}{\micro W} for the effective coupling strength and a value of \unit[$\Gamma_{in} =25.5 \pm 1.2$]{MHz} for the inhomogeneous broadening.

Fig.\,\ref{fig:stats} shows a summary of the linewidth measurements on samples A, B and C. All three samples demonstrate narrow distributions of linewidths, with mean linewidths (m) of \unit[62.1]{MHz}, \unit[62.1]{MHz} and \unit[74.5]{MHz} and standard deviations (SD) of \unit[23.8]{MHz}, \unit[24.6]{MHz} and \unit[50.3]{MHz}, respectively, when assuming log-normal distributions \cite{Kasperczyk2020}. The corresponding standard errors of the mean (SE) are \unit[4.7]{MHz}, \unit[4.3]{MHz} and \unit[8.3]{MHz}, respectively.

A sensible target linewidth is perhaps \unit[100]{MHz}: in a microcavity, the Purcell effect can increase the homogeneous linewidth to several hundred MHz such that an inhomogeneous broadening of about 100 MHz can be tolerated. The empirical cumulative distribution functions (ECDF) show that from the measured distributions, one can extract a 92.3\%, 90.9\%, and 81.1\% chance of measuring a linewidth below \unit[100]{MHz} for sample A, B, and C, respectively. This result emphasizes the excellent optical quality and low charge-noise environment of the created NV centers. Similar probability values are obtained by using a Bayesian approach for analyzing linewidth distributions \cite{Kasperczyk2020}; with this approach, we determine the probability that the next measured line has a linewidth below \unit[100]{MHz} of 92.3\%, 92.1\%, and 78.2\% for sample A, B, and C, respectively. With the same approach we also calculate the probability that the median of the distributions is below this threshold, giving a probability of above 99.9\% for each of the three samples. The higher mean linewidth of the NV centers in sample C could be related to the fact that in this case, the writing was performed deeper inside the diamond where it is particularly difficult to avoid spherical aberrations (even with the corrective capability of the objective lens) such that the vacancies are created over a larger volume compared to the other two samples. Of the 15, 20, and 14 NV centers measured in sample A, B, and C, respectively, 13, 15, and 13 showed clear PLE signals.

\section{Conclusion}
In conclusion, we have demonstrated that a t-SIL enables laser-induced vacancy creation within bulk diamond with pulse energies as low as \unit[5.8]{nJ}. There is a large window in pulse energy before the pulse creates irreversible damage. We interpret this wide range of useful pulse energies as a consequence of working in the tunneling breakdown regime for vacancy creation. The vacancies can be created across the full depth of \SI{40}{\micro\meter} bulk diamond samples. Up to an estimated 92.3\% of the created NV centers have an inhomogeneously broadened linewidth below \unit[100]{MHz}, illustrating that laser writing yields an exceptionally high probability of generating narrow-linewidth NV centers as compared to standard implantation and annealing \cite{Chu2014a,VanDam2019,Kasperczyk2020}. This metric is crucial for applications based on spin-photon entanglement as it is greatly beneficial to perform experiments with minimal pre-selection of NV centers. Importantly, the linewidths presented contain the full inhomogeneous broadening. To the best of our knowledge, this is the lowest charge noise measured to date including the full effects of the off-resonant charge-state repump \cite{Chen2017b,Ruf2019a,Chu2014a}. The narrow linewidths suggest reduced damage in the vicinity of the NV centers compared to other NV creation methods. The t-SIL lowers significantly the threshold pulse energy for vacancy generation. This result points to the feasibility of using a standard Ti:sapphire ultrafast laser without the need for a regenerative amplifier for vacancy creation in diamond.

The main goal of the current work is implementation of the laser-written NV centers into a diamond-cavity system, which in the first stage requires etching of the diamond down to a few micrometers. On account of the strongly reduced lattice damage and improved linewidth statistics compared to other methods, the hope is that after etching the charge noise will remain low. If the low charge-noise is maintained, it should be a key step toward generating spin-spin entanglements at higher rates than what has been achieved so far. In a broader context, the high-optical-quality NV centers created through laser writing present a major advancement not only for cavity applications but also in any application requiring a low charge-noise environment and charge-state-stable color centers.

\section*{Acknowledgement}
We acknowledge financial support from the National Centre of Competence in Research (NCCR) Quantum Science and Technology (QSIT), a competence center funded by the Swiss National Science Foundation (SNF), as well as from the EU FET-OPEN Flagship Project ASTERIQS (grant No. 820394), the SNF project grant No.\ 188521, and the SNF R'Equip grant No.\,170741. TJ acknowledges support from the European Unions Horizon 2020 Research and Innovation Programme under the Marie Sk\l{}odowska-Curie grant agreement No.\,792853 (Hi-FrED) and support from the Polish National Agency for Academic Exchange under Polish Returns 2019 programme (agreement PPN/PPO/2019/1/00045/U/0001). SF acknowledges support from the Initial Training Network (ITN) SpinNANO. We thank Lucas Thiel for skillful coding assistance and Yannik Fontana for reviewing the manuscript.

\bibliography{Paper_LaserWriting_arXiv_Aug2021}

\end{document}